\documentclass[twocolumn]{aastex631}

\usepackage[version=4]{mhchem}
\usepackage{todonotes}
\usepackage{savesym}
\savesymbol{tablenum}
\usepackage[per-mode=symbol]{siunitx}
\DeclareSIUnit{\amu}{amu}
\DeclareSIUnit{\ion}{ion}
\DeclareSIUnit{\ions}{ions}
\DeclareSIUnit{\atper}{at.\%}
\DeclareSIUnit{\bar}{bar}
\restoresymbol{SIX}{tablenum}
\usepackage{hyperref}
\usepackage{url}
\usepackage{amsmath}

\begin{document}
\title{Sputter Yields of the Lunar Surface: Experimental Validation and Numerical Modelling of Solar Wind Sputtering of Apollo 16 Soils}

\shorttitle{Sputtering of the Lunar Surface}
\shortauthors{Brötzner et al.}

\correspondingauthor{Johannes Brötzner}
\email{broetzner@iap.tuwien.ac.at}

\author[0000-0001-9999-9528]{Johannes Brötzner}
\altaffiliation{These authors contributed equally to this work.}

\author[0000-0002-9854-2056]{Herbert Biber}
\altaffiliation{These authors contributed equally to this work.}
\affiliation{Institute of Applied Physics, TU Wien, Wiedner Hauptstraße 8-10/E134, A-1040 Vienna, Austria}

\author[0000-0002-7478-7999]{Paul Stefan Szabo}
\affiliation{Space Sciences Laboratory, University of California, 7 Gauss Way, Berkeley, 94720 CA, USA}

\author[0000-0002-2740-7965]{Noah Jäggi}
\affiliation{Space Science and Planetology, Physics Institute, University of Bern, Sidlerstrasse 5, CH-3012 Bern, Switzerland}
\affiliation{Material Science and Engineering Department, University of Virginia, 395 McCormick Road, Charlottesville, VA 22904, USA}

\author{Lea Fuchs}
\affiliation{Institute of Applied Physics, TU Wien, Wiedner Hauptstraße 8-10/E134, A-1040 Vienna, Austria}

\author[0000-0001-9313-3731]{Andreas Nenning}
\affiliation{Institute of Chemical Technologies and Analytics, TU Wien, Getreidemarkt 9, A-1060 Vienna, Austria}

\author[0000-0002-9405-7889]{Martina Fellinger}
\affiliation{Institute of Applied Physics, TU Wien, Wiedner Hauptstraße 8-10/E134, A-1040 Vienna, Austria}

\author[0000-0003-3172-5736]{Gyula Nagy}
\affiliation{Institute of Applied Physics, TU Wien, Wiedner Hauptstraße 8-10/E134, A-1040 Vienna, Austria}
\affiliation{Department of Physics and Astronomy, Uppsala University, Box 516, SE-752 37 Uppsala, Sweden}

\author[0000-0002-1481-6604]{Eduardo Pitthan}

\author[0000-0002-5815-3742]{Daniel Primetzhofer}
\affiliation{Department of Physics and Astronomy, Uppsala University, Box 516, SE-752 37 Uppsala, Sweden}

\author[0000-0003-0517-6817]{Andreas Mutzke}
\affiliation{Max Planck Institute for Plasma Physics, Wendelsteinstraße 1, DE-17491 Greifswald, Germany}

\author[0000-0001-9451-5440]{Richard Arthur Wilhelm}
\affiliation{Institute of Applied Physics, TU Wien, Wiedner Hauptstraße 8-10/E134, A-1040 Vienna, Austria}

\author[0000-0002-2603-1169]{Peter Wurz}

\author[0000-0003-2425-3793]{André Galli}
\affiliation{Space Science and Planetology, Physics Institute, University of Bern, Sidlerstrasse 5, CH-3012 Bern, Switzerland}

\author[0000-0002-9788-0934]{Friedrich Aumayr}
\affiliation{Institute of Applied Physics, TU Wien, Wiedner Hauptstraße 8-10/E134, A-1040 Vienna, Austria}

\begin{abstract}
Sputtering by solar wind ions is a key process driving the ejection of high-energy particles into the exospheres of airless bodies like asteroids, Mercury and the Moon. 
In view of upcoming missions which will deliver new in-situ data on these exospheres like the Artemis program at the Moon and BepiColombo at Mercury, a deeper understanding of sputtering effects is crucial. 
In this work, we combine sensitive quartz crystal microbalance measurements and numerical simulations to quantify sputter yields of Apollo soil 68501 under solar wind relevant conditions. 
We find that none of the commonly used simulation codes can reliably predict laboratory sputter yields without experimental benchmarks. 
All of the employed packages significantly overestimate the sputter yields of flat samples by at least a factor of 2 for the case of hydrogen. 
When accounting for surface roughness and regolith-like porosity, sputter yields are decreased even further to \qty{7.3e-3}{atoms\per\ion} and \qty{7.6e-2}{atoms\per\ion} for H and He at solar wind energies of \qty{1}{\keV\per\amu}, respectively. 
The reduced yields of porous regolith structures are largely independent of the ion incidence angle, making them applicable across a wide range of lunar latitudes. 
This study highlights the need for experimental validation of sputtering models to ensure accurate predictions for space weathering and lunar exosphere composition.
\end{abstract}


\section{Introduction} \label{sec:intro}
The bombardment of the lunar surface by solar wind ions has been shown to be a process responsible for a variety of space weathering effects. 
These include the formation of nanophase iron particles, the darkening and reddening of reflectance spectra and the amorphisation of rims on mineral grains \citep{pieters_space_2016, hapke_space_2001, loeffler_irradiation_2009}.  
Another consequence of solar wind ion bombardment is the ejection of material through sputtering, which serves as a supply mechanism for species in the lunar exosphere \citep{wurz_particles_2022}. 
Due to the kinetic nature of the sputtering process, the ejected particles carry supra-thermal energies, making ion sputtering a source of high-energy particles in the lunar exosphere, alongside the competing processes of micro-meteoroid impact vaporisation. 

The contribution of ion sputtering to exosphere formation has been explored using self-consistent Monte Carlo models \citep{wurz_monte-carlo_2003, wurz_lunar_2007, hurley_contributions_2017, killen_influence_2022, mura_yearly_2023}. 
These models heavily depend on input data, particularly regarding sputter yields of relevant materials, i.e., the number of atoms ejected per incoming solar wind ion. 
Early experimental efforts to quantify the sputter yields under solar wind ion irradiation date back to the 1960s~\citep{wehner_sputtering_1963, wehner_modification_1963} --- predating NASA's Apollo missions --- with these early studies primarily focusing on metallic samples. 
Experiments with more realistic lunar analogue materials remained rather limited for a long time. 
Consequently, numerical simulations based on the binary collision approximation (BCA), often using the SRIM package \citep{ziegler_srim_2010}, became a primary source for sputter yield data. 

Recent experimental advances have provided new sputter yield data from irradiations of oxide thin films, analogue mineral powders and pressed pellets \citep{szabo_solar_2018, szabo_dynamic_2020,szabo_experimental_2020, biber_sputtering_2022, hijazi_kinetic_2017, hijazi_anorthite_2014, schaible_solar_2017}. 
These pioneering studies already provided some consistent findings.
Firstly, SRIM simulations --- previously considered the standard in this field --- consistently overestimate sputter yields. 
Consequently, the more modern and adaptable SDTrimSP code \citep{mutzke_sdtrimsp_2019} has gained popularity. 
A growing body of literature now explores parameter variations within SDTrimSP to better match experimental results and enhance its predictive capabilities \citep{schaible_solar_2017, szabo_solar_2018, szabo_dynamic_2020, morrissey_solar_2022, jaggi_new_2023, szabo_graphical_2022, jaggi_spubase_2024}.
Secondly, the use of lunar analogue materials like silica and silicate minerals has proven useful in capturing the core physics of the sputtering process. 
\citet{taylor_evaluations_2016} emphasised that simulants can be used in engineering studies, provided their properties match the desired use cases. 
However, a comprehensive investigation of the sputtering behaviour of actual lunar material is still lacking. 
Lastly, many of these experimental studies focused on flat sample films, leaving gaps in understanding how factors like surface roughness and porosity --- crucial for the lunar regolith --- affect sputter yields.

The important influence of this regolith morphology on the sputtering process was already pointed out by \citet{hapke_is_1978}. 
Later, \citet{cassidy_monte_2005} used Monte Carlo simulations to model the influence of sample porosity, roughly reproducing a simple analytical model by \citet{johnson_application_1989}.
\citet{rodriguez-nieva_sputtering_2011} performed molecular dynamics (MD) calculations on porous structures, however, the investigated projectile energies were too high to be applicable to solar wind ion sputtering. 
More recently, \citet{szabo_deducing_2022, szabo_energetic_2023} employed three-dimensional BCA simulations to investigate the reflection of solar wind protons from regolith structures as energetic neutral atoms (ENAs). 
The process of ENA backscattering was furthermore studied by means of Monte Carlo and molecular dynamics simulations by \citet{leblanc_origins_2023} and \citet{verkercke_effects_2023}. 
However, literature on the implications of regolith morphology for sputter yields is still limited.

Lately, \citet{killen_influence_2022} indicated that previous assumptions on sputter yields are inconsistent with MESSENGER MASCS observations of Mercury's exosphere, suggesting that 
sputter yields should instead be lower. 
On a similar note, \citet{nie_lunar_2024} recently used isotope analyses on Apollo soils and came to the conclusion that over geological timescales, micro-meteoroid impact vaporisation is the more effective loss mechanism compared to sputtering also on the lunar surface. 
To accurately model and distinguish between these two processes during exosphere formation, validated physical constraints for each mechanism are essential.

In this work, we present joint experimental and numerical studies on the total, quantitative sputter yields of actual lunar material (Apollo~16 sample 68501) under irradiation with \ce{He} and \ce{H} at a constant energy of \qty{1}{\keV\per\amu} (corresponding to a velocity of approximately \qty{440}{\km\per\s}) rather than relying on lunar analogue or simulant materials. 
We will elucidate the importance of two mechanisms that result in significantly reduced sputter yields, potentially providing an explanation on the suggested predominance \citep{nie_lunar_2024} of micro-meteoroid impact vaporisation over sputtering as a supply mechanism for exospheres: 
the intrinsic overestimation of sputter yields by popular model calculations and a pronounced reduction of the yields attributable to the rough and porous nature of regolith structures. 
As a result, we will provide realistic sputter yield estimates for real lunar regolith. 

\section{Methods} \label{sec:methods}
\subsection{Experimental Methods} \label{subsec:experiments}
\begin{figure}[tbp]
    \centering
    \includegraphics{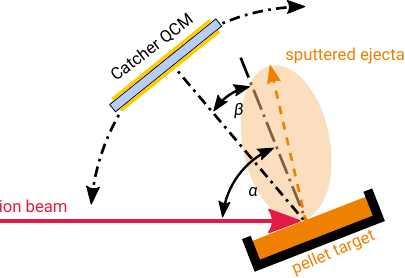}
    \caption{Catcher QCM geometry for measuring the sputter yield. Figure adapted from \citet{biber_sputtering_2022, berger_sputtering_2017}.}
    \label{fig:catcher}
\end{figure}

Sputter yields were measured using a quartz crystal microbalance (QCM) technique in two configurations. 
In the first configuration, a flat, vitreous thin film deposited onto a quartz resonator is used as a sample. 
Using this QCM, mass changes of the film due to ion bombardment are resolved in real time from changes in the quartz resonance frequency, allowing a direct calculation of the sputter yield. 
This setup and technique, described in more detail by \citet{hayderer_highly_1999}, has a resolution down to \qty[per-mode=symbol]{e-11}{\g\per\s} \citep{golczewski_quartz-crystal-microbalance_2009} and has been successfully used in previous studies on analogue materials \citep{szabo_dynamic_2020, szabo_experimental_2020, biber_solar_2020, biber_sputtering_2022}. 
In addition to this common configuration, we applied the catcher QCM method \citep{berger_sputtering_2017, biber_sputtering_2022} where a second QCM is mounted within the experimental vacuum vessel, facing the irradiated sample. 
During ion irradiation, it measures the mass that is deposited onto its surface at the current position. 
This catcher QCM can be rotated with respect to the common centre axis of the setup, subsequently varying the angle $\beta$ between the sample and catcher surface normals (Fig.~\ref{fig:catcher}). 
This extension of the classic QCM setup allows to probe the angular distribution of sputtered ejecta. 
When applying this method to the above described thin film QCMs as samples, both the sputter yields and the corresponding angular distributions of ejecta are measured simultaneously. 
Using these measurements as references, the catcher technique allows to reconstruct sputter yields of bulk samples of the same material, where otherwise no direct information would be available. 
A description of the evaluation is given by \citet{biber_sputtering_2022} while details on the used ion beam setup are given in Appendix~\ref{app:setup}. 

These techniques were applied to two different types of samples prepared from Apollo sample~68501. 
This material is a mature (I\textsubscript{s}/FeO $=85$) soil collected during the Apollo~16 mission with an agglutinate content of about 38\% and average grain sizes of $\approx$\qty{100}{\um} \citep{meyer_lunar_2010}.
We pressed some of the regolith into stainless steel holders to form stable pellets following the procedure outlined by \citet{jaggi_creation_2021}. 
One of these pellets was subsequently used to grow a thin glassy film directly onto a quartz resonator by means of pulsed laser deposition, forming the basis for the QCM experiments. 
From comprehensive analysis we found that the key difference between the deposited films and the pellets on which they are based is found in surface roughness: 
The films are flat while the pellets exhibit a rough surface.
Further details on sample production and characterisation, in particular for surface roughness and chemical composition, are given in Appendix~\ref{app:samples}.

\subsection{Computational Modelling} \label{subsec:simulations}
Simulations of solar wind ion sputtering were carried out using SDTrimSP \citep[version 6.06, ][]{mutzke_sdtrimsp_2019} and SRIM 2013 \citep{ziegler_srim_2010}. 
These codes employ the binary collision approximation that assumes collisions to involve only two particles at a given time, allowing for efficient calculation of energy and momentum transfer between collision partners \citep{eckstein_computer_2007}. 
Subsequently, the impactor and generated recoils are traced on their paths through the sample until their energies fall below a threshold. 
The resulting output includes information on the sputter yield as well as the ejecta angular distributions and energy distributions. 
Compared to SRIM, SDTrimSP allows for variations of more parameters and underlying physical models. 
Details on the chosen settings can be found in Appendix~\ref{app:further_sims}. 
Because the model assumes amorphous samples, these simulated data are directly comparable to the experimental results from the thin film QCM irradiations. 

In addition to the 1D simulations, we applied SDTrimSP-3D \citep[versions 1.21 and 1.22,][]{toussaint_sputtering_2017} to quantify how the surface morphology influences the sputter yield, as roughness and porosity are known to play an important role in the interaction of ions with materials \citep{kustner_influence_1998, cupak_sputter_2021, biber_sputtering_2022, szabo_deducing_2022}. 
We carried out the 3D calculations for the surfaces of the rough pellets as given by atomic force microscopy (AFM) images using the best fitting parameter set from the 1D cases. 
Furthermore, a porous regolith model was implemented as described before by \citet{szabo_deducing_2022}. These SDTrimSP-3D regolith simulations have already been shown to reproduce reflected neutral H spectra measured by Chandrayaan\nobreakdash-1, which strongly indicates an accurate description of the ion-regolith interaction in the model. 
Using this approach, the influences of roughness and porosity on the sputter yield compared to a flat surface can be untangled. 

SDTrimSP and the adapted binding energy model from \citet{jaggi_new_2023} also form the basis for SpuBase~\citep{jaggi_spubase_2024, jaggi_nvjaeggispubase_2024}, a database of already carried out simulations for flat surface samples. 
SpuBase offers sputter yields, ejecta angular and energy distributions for a wide variety of minerals. 
Results for bulk samples of a given atomic composition are then superposed from the constituent minerals. 

\section{Results} \label{sec:results}
\subsection{Flat Sample Sputter Yields} \label{subsec:flat_yields}
\begin{figure*}[htbp]
    \centering
    \includegraphics{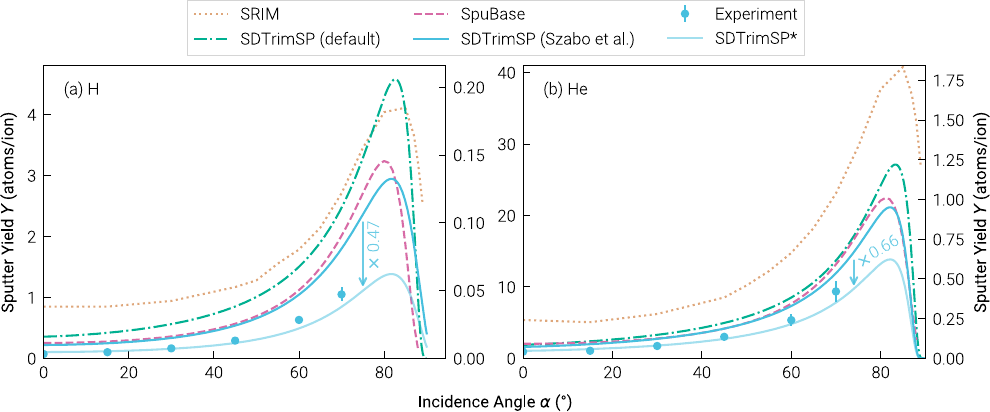}
    \caption{Sputter yields for flat surfaces. Comparison of various simulation models (lines) and experimental results (points) for \qty{1}{\keV\per\amu} H (a) and He (b) impactors, respectively. 
    Simulation data stem from SRIM (dotted line), SpuBase (dashed), SDTrimSP with the default parameters (dashed-dotted) and with the parameters proposed by \citet{szabo_dynamic_2020} (solid). 
    Additionally, the arrows quantify the offset between experiment and simulation, and the lighter coloured lines give the simulations scaled to match experimental data. 
    The asterisk in ``SDTrimSP*'' denotes the use of the parameter set proposed by \citet{szabo_dynamic_2020} combined with this scaling factor.}
    \label{fig:flat_yields}
\end{figure*}

Laboratory sputter yields of flat samples are given in Figure~\ref{fig:flat_yields} alongside simulation data of common modelling approaches. 
The sputter yields are given as function of incidence angle $\alpha$ in atomic mass units per incident ion (left axis) and atoms per incident ion under the assumption of stoichiometric particle fluxes (right axis). 

For the H case, it is clear that all modelling attempts overestimate the experimental results denoted by the filled circles. 
SRIM (dashed line) predicts the highest values throughout most incidence angles. 
SDTrimSP with its default parameter set (dashed dotted line) is lower, with an exception in the narrow region around $\alpha=\ang{80}$. 
The two lines resulting in the lowest total mass yields are SDTrimSP using the parameter set published by \citet{szabo_dynamic_2020} (solid) and SpuBase with an improved hybrid compound binding model (dashed). 
While for smaller incidence angles, the two give almost identical results, the SpuBase curve rises steeper beyond $\alpha \approx \ang{70}$ and reproduces the experimental trend better. 
SpuBase, however, cannot be extended to three-dimensional studies. 
We therefore proceeded to use the model by \citet{szabo_dynamic_2020} also for the subsequent 3D investigations and quantified the discrepancy between the solid line in Fig.~\ref{fig:flat_yields} and the experimental data to be a factor of $0.47$. 
This scaling factor was determined by taking the point-wise ratio between numerical and experimental data points and subsequently averaging these fractions. 

A similar situation is observed for He. 
However, in this case a clear separation between SRIM and the SDTrimSP variants is observed and SRIM gives the highest sputter yield values over all incidence angles. 
Once more, SpuBase is the SDTrimSP-based curve without any adaptation closest to the measured data. 
On the other hand, the \citet{szabo_dynamic_2020} approach results in the lowest mass yields. 
In this case, a factor of $0.66$ is necessary to match this curve to the experimental data points. 
While for He, the overestimation by SDTrimSP-based models is moderate, it is notably worse for H, the most prominent solar wind species, across all investigated numerical approaches. 

\subsection{Surface Morphology Dependent Sputter Yields} \label{subsec:3d_yields}

\begin{figure}[tbp]
    \centering
    \includegraphics{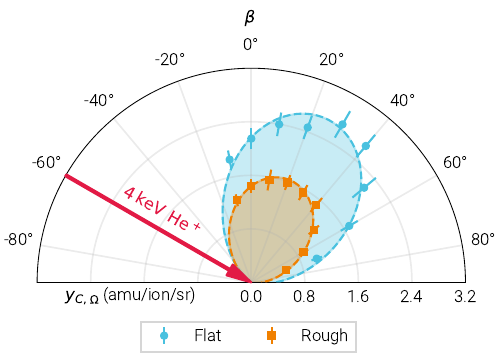}
    \caption{Sputtered particles polar angle distributions as measured by the catcher QCM. 
    Mass deposited onto the catcher QCM $y_{C, \Omega}$ in atomic mass units normalised per incidence ions and detector solid angle as a function of detector position $\beta$ (Fig.~\ref{fig:catcher}). 
    Experimental data (symbols) were fitted using a modified cosine \citep[cf.][]{biber_sputtering_2022} and the difference in sputter yield is represented by the different integrals (shaded areas).} 
    \label{fig:catcher_results}
\end{figure}

To experimentally derive sputter yields for the rough pressed pellets, the catcher QCM method described in Section~\ref{subsec:experiments} was employed. 
During this process, the ejecta angular distributions are probed in addition to the absolute mass yields. 
One representative measurement comparing the emission characteristics of the flat and rough samples under \qty{1}{\keV\per\amu} He impinging at an incidence angle of \ang{60} is shown in Figure~\ref{fig:catcher_results}. 
The evaluation of the sputter yields follows \citet{biber_sputtering_2022}, and the difference in sputter yield for the different samples is clearly visible. 

\begin{figure*}[hbt]
    \centering
    \includegraphics{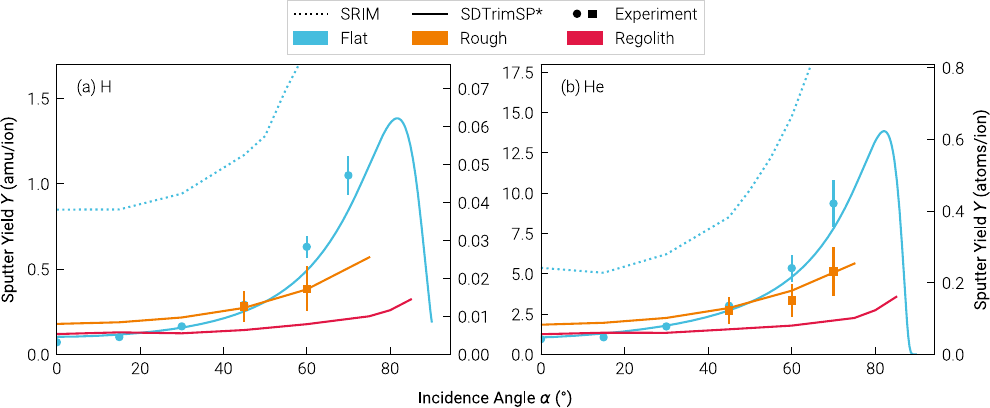}
    \caption{Sputter yields of lunar surface material for typical solar wind impactors and different surface morphologies comprised of a 
    flat surface (blue), a rough pressed pellet surface (orange) and a porous regolith structure (red). 
    Experimental results are denoted by the symbols with error bars, while SDTrimSP-based and SRIM simulations are given by solid and dotted lines, respectively. 
    Note that the scaling factor derived in Figure~\ref{fig:flat_yields} has been applied to the SDTrimSP simulations, as indicated by the labelling convention using the asterisk.} 
    \label{fig:yields}

\end{figure*}

The total sputter yields for all samples are given in Figures~\ref{fig:yields}(a) and \ref{fig:yields}(b) for \qty{1}{\keV\per\amu} \ce{H} and \ce{He}, respectively. 
Note that for SDTrimSP-based curves (both 1D and 3D), the projectile-dependent scaling factors of 0.47 and 0.66 (Fig.~\ref{fig:flat_yields}) are applied. 

Both ion species show a typical dependence of the sputter yield on the incidence angle when flat samples are considered (blue symbols and lines in Figure~\ref{fig:yields}). 
The yield increases with $\alpha$ until a maximum is reached for grazing incidence at roughly \ang{80} for both \ce{He} and \ce{H}. 
Beyond this point, a sharp decrease is reported and the sputter yield approaches 0 for near-horizontal incidence. 
Data for the rough pellet sample (orange lines) show a reduction in sputter yield for large incidence angles and an increase compared to the flat sputter yields (blue lines) for near-normal impact.
The crossing point where sputter yields coincide between flat and rough samples is approximately located at $\alpha = \ang{45}$ for both ion species. 

When porosity is introduced through the implementation of regolith structures in SDTrimSP-3D (red lines), the same effect is observed in an even more pronounced way; 
the sputter yield is further flattened and reduced. 
In this case, equal yields as compared to a flat surface are achieved near $\alpha = \ang{20}$, but until this point neither curve exhibits a significant slope such that also at normal incidence, the sputter yields are comparable. 
Beyond this region, the regolith yield stays almost constant and only for grazing incidence above \ang{75} to \ang{80} a slight increase is discernible. 
When averaged over the simulated angle range, the regolith sputter yields for lunar soil are \qty{7.27e-3}{atoms\per\ion} for H and \qty{7.62e-2}{atoms\per\ion} for He.

\section{Discussion} \label{sec:discussion}
\subsection{Implications for Simulations} \label{sec:disc_simulations}
BCA simulations have been used for decades and are still an active topic in research and development \citep{hofsass_binary_2022, jaggi_new_2023, hofsass_low_2023}. 
Nonetheless, common modelling approaches struggle with correctly describing sputter yields for compound materials (Fig.~\ref{fig:flat_yields}).
It is evident that SRIM cannot reliably predict sputter yields for the lunar surface. 
This was to be expected, as the shortcomings of SRIM have been known and its use, particularly for energies in the solar wind relevant regime, is discouraged \citep{wittmaack_reliability_2004, wittmaack_misconceptions_2016, hofsass_simulation_2014,shulga_simulation_2024, shulga_note_2018}. 
Also, the overestimation of the sputter yield by SRIM for mineral samples has been shown repeatedly \citep[e.g.,][]{szabo_solar_2018, szabo_dynamic_2020, schaible_solar_2017}. 
Nevertheless, SRIM is still used in recent literature \citep[see, for example, publications by][]{rubino_space-weathering_2024, nenon_long-term_2020, leblanc_origins_2023}. 
As an alternative, SDTrimSP has been suggested, as its predictions are in better agreement with experimental results \citep{szabo_dynamic_2020, szabo_solar_2018, schaible_solar_2017, biber_solar_2020, biber_sputtering_2022}. 
However, also for this code, parameter adaptations (or scaling) are necessary to reproduce sputter yields measured in experiments. 

One possible shortcoming hindering a better understanding of the sputtering process in the BCA picture is the knowledge gap concerning surface binding energies (SBEs).
The SBE directly influences the sputter yield \citep{sigmund_theory_1969, sigmund_recollections_2012} and is often approximated as the energy of sublimation \citep{kelly_surface_1986, gnaser_energy_2007}. 
This view, however, is debated and subsequently, a substantial amount of research has been carried out on the physical meaning of the SBE and its significance for sputtering, also in the space sciences context \citep{szabo_dynamic_2020, morrissey_sputtering_2021, morrissey_solar_2022, killen_influence_2022, hofsass_binary_2022, jaggi_new_2023}. 
As the SBE is strongly related to the energy spectra of sputtered ejecta \citep{thompson_ii_1968}, measurements of ejecta energy distributions could potentially clarify these matters and improve BCA simulations. 
However, few data are currently available for relevant compound materials, and only some data sets have been published for metallic samples or alkali halides \citep{husinsky_application_1985, wucher_energy_1986, betz_energy_1994, dukes_lunar_2015}. 
For the time being, the \citet{jaggi_new_2023} model used in SpuBase \citep{jaggi_spubase_2024} is the one studied BCA model closest to the experiment where the binding energy approach is physically motivated and does not stem from fitting to a data set. 
Nonetheless, the remaining uncertainties in the BCA codes underline the necessity to validate simulated sputter yields with available experimentally measured data sets.

\subsection{Sputter Yield Reduction through Surface Morphology} \label{subsec:disc_yields}
In addition to the overestimation by simulations, surface morphology reduces the effective sputter yields (Fig.~\ref{fig:yields}). 
\citet{biber_sputtering_2022} found a similar sputter yield reduction when comparing enstatite (\ce{MgSiO3}) thin film and pressed pellet samples and ascribed this effect to surface roughness. 
Indeed, \citet{cupak_sputter_2021} demonstrated in a joint experimental and numerical study that surface morphology is capable of lowering the sputter yield and highlighted a correlation between this decrease and the mean of the surface inclination angle distribution. 
Furthermore, this result was reinforced by an analytical investigation arriving at the same conclusion \citep{szabo_analytical_2022}. 

In this work, the pellet roughness (quantification in Appendix~\ref{app:samples}) is comparable to the one in \citet{biber_sputtering_2022}, as are the measured sputter yield reductions. 
In a further step, \citet{cupak_sputter_2021} provide a Monte-Carlo-style code called SPRAY allowing to calculate sputter yields from AFM images of a given sample. 
In this case, angle and energy dependent simulation results from BCA codes are mapped onto the input surface by a ray-tracing algorithm. 
That way, any deviations from flat surface sputter yields are unambiguously attributable to surface morphology, as no other parameter is varied. 
We found excellent agreement to our experimental results using this code as well, pointing towards surface roughness as the main driver behind the observed sputter yield reduction for the pellet samples. 
A comparison of the SPRAY results to laboratory data and SDTrimSP is presented in Appendix~\ref{app:spray}. 
Additionally, the observed matching sputter yields for $\alpha = \ang{45}$ and the reduction to about half of the thin film sputter yield at \ang{60} match well with the analytical predictions by \citet{szabo_analytical_2022} for the given roughness. 

In contrast to these morphology effects, we do not expect crystallinity to play a role in the sputter yield modifications. 
While it is known that crystal structure has an effect on the sputtering properties of a sample \citep[see, e.g.,][]{onderdelinden_influence_1966} and that the sputtering behaviour of amorphous and polycrystalline samples is not necessarily the same \citep{schlueter_absence_2020}, these considerations are irrelevant in the context of this study:
Our flat samples are amorphous by the nature of their production process \citep[as shown by X-ray diffraction in][]{szabo_solar_2018, szabo_experimental_2020}. 
The pellets were pressed not from pristine minerals, but rather from regolith that naturally expresses amorphous rims around its crystalline sample fraction. 
Although fresh surfaces might have been created by breaking grains during the pellet pressing process, a \qty{4}{\keV} He fluence of \qty[per-mode=reciprocal]{7.31e17}{\per\cm\squared} was applied to the samples during the first preparatory irradiation. 
\citet{carrez_low-energy_2002} showed that a \qty{4}{\keV} \ce{He} fluence of \qty[per-mode=reciprocal]{5E16}{\per\cm\squared} is sufficient to amorphise a rim of olivine (\ce{(Mg,Fe)2SiO4}) with a resulting thickness of several \qty{10}{\nm}.
We therefore deem both the thin film and the pellet experimental results comparable to the (amorphous) BCA simulations both in the 1D and 3D configuration and suitable to benchmark these very numerical results. 
Sputter yield data for the rough pellets match excellently between experiments and SDTrimSP-3D simulations after the application of the correction factor determined in Figure~\ref{fig:flat_yields} from flat sample data. 
It is thus justified to apply the same procedure to the porous regolith structures. 
Consequently, these regolith sputter yields should be used as realistic supply rates when modelling the lunar exosphere formation by solar wind ion impact.

\subsection{Comparison to Previous Measurements of Single Minerals}
\begin{figure*}[t]
    \centering
    \includegraphics{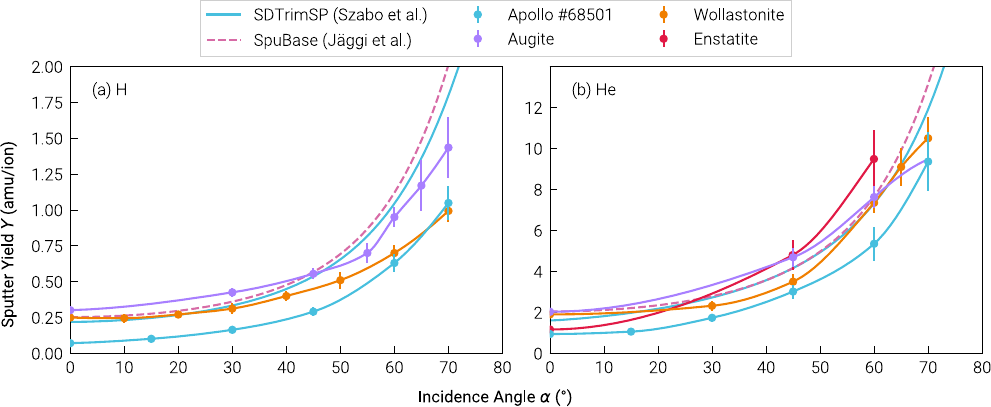}
    \caption{Compilation of experimentally measured sputter yields, both from this study and previous investigations using analogues, compared to two approaches of SDTrimSP simulations applied to the composition of the Apollo~68501 sample \citep{szabo_dynamic_2020,jaggi_spubase_2024}. 
    Connecting lines between experimental data are interpolated to guide the eye. } 
    \label{fig:mineral_comparison}
\end{figure*}

In the past years, various studies reported sputter yields for lunar analogue materials that were measured using the same method as described in this paper, where mass changes of flat, amorphous films were resolved by means of a QCM. 
Figure~\ref{fig:mineral_comparison} compares the Apollo~68501 laboratory sputter yields from this work to data from wollastonite~\citep[\ce{CaSiO3,}][]{szabo_solar_2018, szabo_dynamic_2020}, augite~\citep[\ce{(Ca,Mg,Fe)2Si2O6},][]{biber_influence_2023} and enstatite~\citep[\ce{MgSiO3,}][]{biber_sputtering_2022}. 
In addition, the yields predicted by both SpuBase and SDTrimSP using the \citet{szabo_dynamic_2020} parameters without scaling are shown. 
Both the regolith and the mineral samples share a similar composition with O and Si abundances of roughly \qty{60}{\atper} and \qty{20}{\atper}, respectively. 
As the sputter yield in the equilibrium is governed by the bulk stoichiometry, one would expect similar total mass yields across these samples with the differences arriving from the variation of the metal species. 
This is the case and most of the single-mineral data points lie between our Apollo data and the model predictions for the regolith composition. 

Note that the yields measured from the lunar material are the lowest across all numerical and laboratory data sets (Fig.~\ref{fig:mineral_comparison}).
While the models work well for individual minerals \citep[see, e.g., Figure 4 in][]{jaggi_new_2023}, they overestimate data for the case of the more complex Apollo soil. 
Sample roughness and composition cannot be the reason for this discrepancy, as these parameters are well controlled experimentally across all these studies and comparable to the simulated cases. 
Moreover, we also varied the composition input to the simulations within the error bars from the sample analysis (Table~\ref{tab:compo}) and found only an insignificant level of deviation in the results. 
A more likely explanation could be the formation of bonds that exceed the ones typically found within an amorphous silicate layer. 
For example, both the \cite{jaggi_new_2023} model and the \cite{szabo_dynamic_2020} approach reach the best agreement with laboratory data assuming the importance of oxygen in the bond structure --- either considering the oxide formation energy or by directly increasing the oxygen binding energy followed by averaging of all the binding energies. 
This neglects bonds that could form between species of the different minerals. 
Should those bonds be stronger than the ones found in the bulk material, then the resulting higher binding energy would provide an explanation for the lower sputter yields. 
Another way of reducing yields would be by decreasing the target density and consequently increasing the binary collision mean free path.
It is unclear why these effects are not found in glassy thin films of single mineral analogues. 
We propose that the high number of components in the Apollo sample would favour the formation of either longer (lower density) or stronger bonds (higher BEs) than are found in its components.  
This is mirrored in the way SpuBase handles such complex materials: 
Rather than assuming a glass of homogeneous composition (as is the case for the film on the QCMs and the \cite{szabo_dynamic_2020} SDTrimSP approach), they are decomposed from constituent minerals. 
While the difference between experiments and models underlines the necessity for experimental validation, particularly for complex samples, the proposed explanation can be tested by measurements of ejecta energy distributions of both the Apollo samples and the individual minerals. 
In addition to the arguments in Section~\ref{sec:disc_simulations}, this highlights once more the necessity for laboratory studies on sputtered ejecta energy distributions.

\section{Summary and Conclusion} \label{sec:conclusion}
For the first time an advanced quartz crystal microbalance setup was used to measure sputter yields of returned lunar material (Apollo~16 sample 680501) under solar wind-like conditions. 
Two types of samples were prepared: 
flat stoichiometric films to study the incidence angle dependence of the sputter yield and rough pellets from pressed regolith to examine the effect of surface morphology. 
Our results show that all commonly used binary collision approximation models overestimate the sputter yield for flat surfaces, albeit the SpuBase model \citep{jaggi_new_2023,jaggi_nvjaeggispubase_2024} came closest to experiments \emph{ab initio}. 
The rough pellets from pressed regolith were used to study the influence of surface roughness on the sputter yield and used to validate the predictions of more advanced 3D-BCA codes. 
We then implemented a three-dimensional regolith model with a proven track record that more accurately replicates lunar surface conditions. 
Advanced BCA simulations using this regolith model revealed that the porosity further reduces the sputter yield of lunar soil and that the resulting yields are largely independent of the ion incidence angle, corresponding to the solar zenith angle on a macroscopic scale.
The realistic porous regolith sputter yields were found to be \qty{7.3e-3}{atoms\per\ion} for H and \qty{7.6e-2}{atoms\per\ion} for He at solar wind energies of \qty{1}{\keV\per\amu}.

The present and upcoming space missions to the Moon (e.g., the Artemis program and Mercury (BepiColombo, \citet{orsini_serena_2021}) will provide highly resolved in-situ data of the surface and exospheric compositions, including more details about their spatial and temporal dynamics. 
Our study results provide constraints on the actual source rates for sputtering, potentially aiding the interpretation of recent and future space measurements.

\section*{Data Availability}
Data presented in this article is publicly available at \url{https://doi.org/10.48436/zkkzn-x6j34} under a CC BY 4.0 licence. The data set includes more ejecta angular distributions than shown in Fig~\ref{fig:catcher_results}.

\begin{acknowledgments}
This research was funded in whole or in part by the Austrian Science Fund (FWF) [\url{https://doi.org/10.55776/I4101}]. 
For open access purposes, the author has applied a CC BY public copyright license to any author accepted manuscript version arising from this submission.
Funding was also provided by the Swiss National Science Foundation Fund (200021L\_182771/1, P500PT\_217998). 
J.B. acknowledges financial support by KKKÖ of ÖAW. 
The authors gratefully acknowledge support from NASA’s Solar System Exploration Research Virtual Institute (SSERVI) via the LEADER team, grant No. 80NSSC20M0060. 
We are very grateful to NASA for providing a lunar regolith sample collected during the Apollo~16 mission via the Lunar Sample Request program (\url{https://curator.jsc.nasa.gov/lunar/sampreq/requests.cfm}). 
Ion beam analysis of the lunar material at UU was supported by the RADIATE project under the Grant Agreement 824096 from the EU Research and Innovation program HORIZON 2020. 
Accelerator operation at Uppsala University is supported by the Swedish Research Council VR-RFI (contract \#2019\_00191). 
The computational results presented have been achieved [in part] using the Vienna Scientific Cluster (VSC). 
J.B. acknowledges the pylustrator software~\citep{gerum_pylustrator_2020}.
\end{acknowledgments}

\appendix
\restartappendixnumbering
\section{Irradiation setup} \label{app:setup}
The ion beam irradiation setup is the same as in \citet{szabo_solar_2018} and \citet{biber_sputtering_2022}. 
It consists of a \qty{14.5}{\GHz} electron cyclotron resonance ion source and an \(m/q\) separation achieved via a magnetic sector field \citep{galutschek_compact_2007}. 
A set of computer-controlled deflection plates in front of the first aperture is used for switching the ion beam on/off electronically without moving parts, to minimise interference with the sensitive QCM signal. 
Scanning plates are used to ensure homogeneous sample irradiation.
Furthermore, a Prevac FS40A1 electron flood source (up to \qty{100}{\uA} low energy electrons, $<$~\qty{20}{\eV}) was used to prevent charging of the insulating pellets due to the impinging ion beam. 
The \qty{1}{\keV\per\amu} H data were obtained from double-energy \ce{H2} irradiations. 
This is a common practice, and the underlying assumption \citep[e.g.,][]{szabo_solar_2018} is that a \qty{2}{\keV} hydrogen molecule is dissociated at the surface and acts as two independent hydrogen atoms of \qty{1}{\keV} each. 
In the presented energy regime where sputtering is dominated by linear collision cascades, no effects are expected to arise from the molecular structure of the projectiles. 
Nonlinearities occur at much lower energies~\citep{dobes_sputtering_2011}, or considerable higher ones~\citep{andersen_heavy-ion_1975}. 
Moreover, this has been experimentally verified~\citep{kenknight_sputtering_1964, szabo_experimental_2020} and numerically checked by means of MD simulations~\citep{brotzner_sputtering_2023}. 
Typically achieved ion fluxes are in the order of \qty[per-mode=reciprocal]{3E12}{\per\cm\squared\per\s} for \ce{He+} and \qty[per-mode=reciprocal]{1e13}{\per\cm\squared\per\s} for \ce{H2+}.

\section{Sample Characterisation} \label{app:samples}
\begin{figure*}[htbp]
    \centering
    \includegraphics{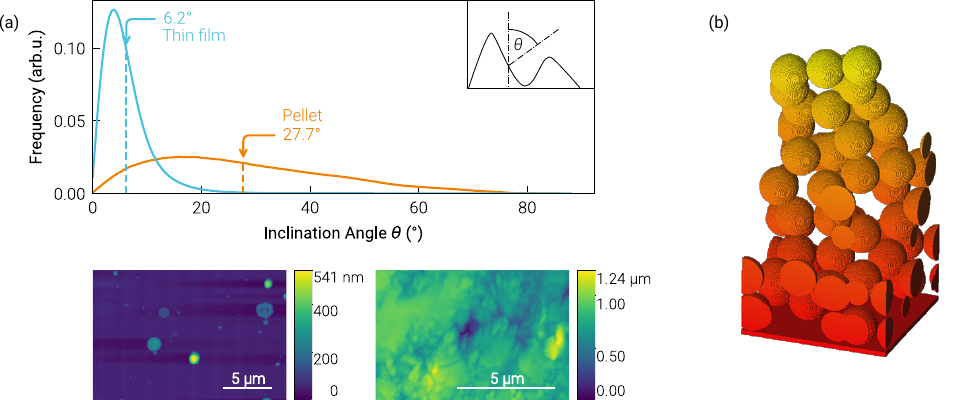}
    \caption{Visualisation of sample properties used during this study. 
    (a) AFM images of the thin film sample (left image in the bottom row) and the pressed pellet (right image in the bottom row). 
    The top panel shows the surface inclination angle distribution for both experimentally used sample types, thin film (blue) and pellet (orange). 
    The means of the distributions are marked by the respective vertical dashed lines and arrows. 
    The inset defines the local inclination angle $\theta$ as the angle between the local and the global surface normal. 
    (b) Example of the regolith structure with a porosity of about 0.8 \citep[see][]{szabo_deducing_2022}, which was used to simulate the sputter yield for the regolith case.}
    \label{fig:siad_afm}
\end{figure*}

All samples used in this study were produced from Apollo soil 68051.
Pellets were pressed following the procedure in \citet{jaggi_creation_2021}, but without filtering for grain sizes. 
A layer of KBr was first pressed into the holder to increase cohesion between the regolith and the back-plate. 
Thin films were deposited onto QCMs by means of pulsed laser deposition. 
This was done under an \ce{O2} atmosphere of \qty{4e-2}{\milli\bar} to achieve stoichiometric oxygen concentration in the resulting film.
A \ce{KrF} excimer laser was used with a wavelength of \qty{248}{\nm} and a pulse frequency of \qty{5}{\Hz} at a pulse energy of \qty{400}{\m\J\per pulse}.

The chemical composition of both sample types was analyzed using a combination of ion beam analysis techniques: time-of-flight elastic recoil detection analysis (ToF-ERDA), Rutherford backscattering spectrometry (RBS) and particle-induced X-ray emission (PIXE). 
Details of the experimental set-ups can be found in \citet{strom_ion_2022}. 
ToF-ERDA was carried out using a combined anode gas ionisation chamber and time-of-flight as detector with a primary beam of \qty{36}{\MeV} \ce{^127I^{8+}} and an incident angle of \ang{67.5}. 
To unambiguously distinguish signals from species with similar atomic masses (i.e., Al and Si, K and Ca, Cr and Fe), RBS simultaneously to PIXE was performed using He at 2 and 5.5 MeV as primary beam at normal incidence.
While RBS provides accurate quantification of Al and Si, PIXE is able to detect the presence of K and Cr in significantly smaller amounts compared to Ca and Fe, respectively. Finally, µ-beam RBS/PIXE was used to verify the lateral homogeneity of the samples using \qty{4}{\MeV} He as primary beam \citep{nagy_scanning_2022}.
Different regions across the entire samples (including centre and edge regions) were analysed using a beam spot of 4-5 microns (\qtyproduct{1 x 1}{\mm} area per analysis). 
Results indicate that the composition of the pellet and thin film is homogeneously distributed along the samples.
The atomic concentration of the main components observed in the samples are presented in Table~\ref{tab:compo} and agree well with literature \citep{meyer_lunar_2010, bansal_chemical_1972}.

\begin{deluxetable}{lccc}[htbp]
\tablecaption{Sample Compositions\label{tab:compo}}
\tablehead{\colhead{} & \colhead{Thin film} & \colhead{Pellet} & \colhead{Literature}}
\startdata
O  & $61.0 \pm 0.6$ & $58.6 \pm 0.6$ &61.0\\
Si & $14.3\pm0.4$   & $17.0\pm0.5$   &16.3\\
Al & $9.96\pm0.4$   & $11.9\pm0.5$   &11.3\\
Ca & $8.14\pm0.2$   & $7.26\pm0.2$   &5.93\\
Mg & $3.27\pm0.1$   & $3.1\pm0.1$    &3.37\\
Fe & $2.76\pm0.1$   & $1.47\pm0.1$   &1.65\\
Ti & $0.45\pm0.1$   & $0.18\pm0.1$   &0.16
\enddata
\tablecomments{Concentrations of the main components of the samples in \unit{\atper} obtained by combining ToF-ERDA, RBS and PIXE. For comparison, data from \citet{meyer_lunar_2010} after \citet{bansal_chemical_1972} are given.}
\end{deluxetable}

In addition to the chemical analysis, AFM images were taken to characterise the surface roughness of both the thin film and pellet samples. 
No change in surface roughness was found after the ion beam experiments. 
The surface roughness was quantified using the surface inclination angle distribution method proposed by \citet[][]{cupak_sputter_2021}, the results of which are presented in Figure~\ref{fig:siad_afm}. 
The bottom row in Figure~\ref{fig:siad_afm}(a) shows AFM images of the thin film (left) and the pellet sample (right) and illustrates the difference in roughness:
The film sample is generally flat with the exception of some particles that formed during the PLD deposition. 
This difference in roughness is quantified by the respective surface inclination angle distribution and their means in the top panel, where the blue and orange lines denote the thin film and pellet, respectively. 

Figure~\ref{fig:siad_afm}(b) shows a model of the porous regolith-structured sample used for the simulations. 
These structures were created using the parameters best fitting to reproduce ENA emission from backscattered solar wind H as described in \citet{szabo_deducing_2022, szabo_energetic_2023}. 
The porosity is accordingly defined as the fraction of empty space between the regolith grains and the volume of the simulation cell from its lower boundary to the topmost grain.

\section{Simulation details} \label{app:further_sims}
Four numerical models based on the BCA were studied. 
SRIM simulations were carried out using the damage calculation model ``Detailed Calculation with full Damage Cascades''. 
Apart from that, no other adaptations were made. 

For SDTrimSP with its default settings, simulations were run dynamically, accounting for fluence dependent changes in surface composition. 
This is advocated for by \citet{morrissey_establishing_2023}. 
Other recommendations from this work include the use of a mixed ion beam (96\% H, 4\% He) as well as using distributions for incidence angles and energies. 
Due to the angle resolved comparison to experimental data obtained from monatomic projectiles of a well defined energy, we did not apply these. 
Finally, they propose to use adapted SBEs, if known. 
Such are however not available to our knowledge for the studied sample. 

In the approach published by \citet{szabo_dynamic_2020}, the surface binding model in SDTrimSP is set to \texttt{isbv $=2$}, thereby averaging the SBEs of the constituent species. 
The oxygen SBE is increased to \qty{6.5}{\eV} and the sample density is set to reflect the actual material density, in our case \qty{3.1}{\g\per\cm\cubed} after \citet{heiken_lunar_1991}.

Finally, in SpuBase the simulations are already carried out for 21 rock-forming minerals into which the sample is decomposed.
The improved hybrid binding model for compounds (HB-C) from \citet{jaggi_new_2023} is used in the calculations for the individual minerals, offering an improvement over arbitrary adaptation. 
However, data are only available for the 1D case and effects of surface morphology cannot be studied. 
The HB-C model also cannot be used directly in standalone SDTrimSP simulations as the high number of components used in this study is not supported.

\section{Comparison with ray-tracing algorithm SPRAY} \label{app:spray}
\begin{figure}[htp]
    \centering
    \includegraphics{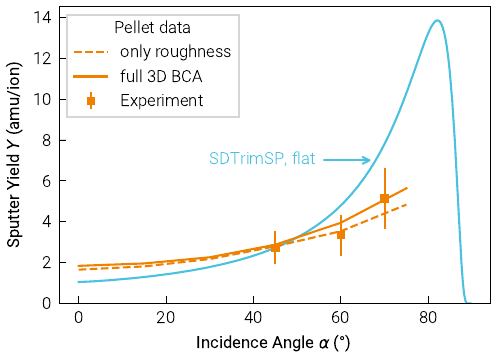}
    \caption{Comparison of simulation results for the rough pellet sample under \qty{1}{\keV\per\amu} \ce{He} bombardment: SDTrimSP-3D, which calculates full collision cascades using the BCA, is given by the solid orange line, while SPRAY (only mapping 1D results onto the sample surface, see Section~\ref{subsec:disc_yields}) is represented by the dashed orange line. Both match within the error bars with our experiments (orange squares). 1D SDTrimSP (blue line) is shown for reference.}
    \label{fig:spray_comp}
\end{figure}

We performed simulations using the SPRAY algorithm presented by \citet{cupak_sputter_2021} to elucidate the origin of the sputter yield reduction observed in Figure~\ref{fig:yields}. 
It uses ray tracing to map results from 1D BCA simulations onto triangulated surfaces that stem from microscopy images and inherently considers the variation of the local impact angle and surface shadowing while also taking into account redeposition and secondary sputtering by reflected ions. 
Due to this nature, any arising deviations from flat surface sputter yields originate from the surface morphology. 
Results are shown in Figure~\ref{fig:spray_comp} by the dashed orange line and agree excellently with experimental findings. 
They are within the experimental error bars indistinguishable from the full 3D BCA simulations presented by the solid orange line, corroborating the discussion in Section~\ref{subsec:disc_yields}.

\bibliography{references}{}
\bibliographystyle{aasjournal}
\end{document}